\RequirePackage{fix-cm}
\documentclass[twocolumn]{svjour3}          
\smartqed  
\usepackage{graphicx}
\usepackage{natbib}
\usepackage{marvosym}
\usepackage{enumerate}
\usepackage{amssymb}
\usepackage{wasysym}
\usepackage{longtable}
\usepackage[breaklinks]{hyperref}

\def\mathbi#1{\textbf{\em #1}}
\newcommand{\be}{\begin{equation}}
\newcommand{\ee}{\end{equation}}
\newcommand{\fig}[1]{Fig.~\ref{fig_#1}}

\renewcommand{\a}{{\mathbi a}}
\newcommand{\rmax}{R_0}
\newcommand{\tc}{T_{\rm c}}
\newcommand{\tl}{T_{\rm 1/2}}
\newcommand{\deltap}{\Delta p}
\newcommand{\del}{\mbox{\boldmath$\nabla$}}
\newcommand{\sym}{$\bullet$}

\journalname{Exp Fluids}

\begin{document}

\title{The Quest for the Most Spherical Bubble\thanks{This research was supported by the Swiss NSF (Grant No.~200020-116641 and PBELP2-130895) and the European Space Agency ESA.}}

\subtitle{Experimental Setup and Data Overview}

\titlerunning{The Quest for the Most Spherical Bubble} 

\author{Danail Obreschkow  \and
        Marc Tinguely \and
        Nicolas Dorsaz \and
        Philippe Kobel \and
        Aurele de Bosset \and
        Mohamed Farhat
}

\authorrunning{Obreschkow \textit{et al.}, 2013} 

\institute{
	D. Obreschkow (\Letter)\at
	International Centre for Radio Astronomy Research (ICRAR)\\
	University of Western Australia,
	35 Stirling Hwy, Crawley, WA 6009, Australia
	\email{danail.obreschkow@uwa.edu.au}
	\and
	M. Tinguely, P. Kobel, A. de Bosset, M. Farhat \at
	Laboratoire des Machines Hydrauliques \\
	EPFL, 1007 Lausanne, Switzerland
	\and
	N. Dorsaz \at
	Department of Chemistry \\
	University of Cambridge, Cambridge CB2 1EW, UK
}

\date{Received: date / Accepted: date}

\maketitle

\begin{abstract}
We describe a recently realized experiment producing the most spherical cavitation bubbles today. The bubbles grow inside a liquid from a point-plasma generated by a nanosecond laser pulse. Unlike in previous studies, the laser is focussed by a parabolic mirror, resulting in a plasma of unprecedented symmetry. The ensuing bubbles are sufficiently spherical that the hydrostatic pressure gradient caused by gravity becomes the dominant source of asymmetry in the collapse and rebound of the cavitation bubbles. To avoid this natural source of asymmetry, the whole experiment is therefore performed in microgravity conditions (ESA, 53rd and 56th parabolic flight campaign). Cavitation bubbles were observed in microgravity ($\sim\!0g$), where their collapse and rebound remain spherical, and in normal gravity ($1g$) to hyper-gravity ($1.8g$), where a gravity-driven jet appears. Here, we describe the experimental setup and technical results, and overview the science data. A selection of high-quality shadowgraphy movies and time-resolved pressure data is published online.
\end{abstract}

\section{Introduction: A roadmap towards unification}\label{section_intro}

Cavitation bubbles remain a key topic in fluid dynamics. While traditionally associated with erosion damage \citep{Philipp1998}, they are currently reconsidered in a wide range of modern applications within food technology \citep{Mason1996}, water cleaning \citep{Wolfrum2003}, medicine \citep{Leighton2010}, and microfluidics \citep{Tandiono2011}. This extraordinary breadth of applications originates from the rich phy\-sics governing the collapse of individual bubbles. Fundamental science is now asked to provide a robust framework unifying this richness.

\begin{figure*}[t]
\centering
  \includegraphics[width=0.8\textwidth]{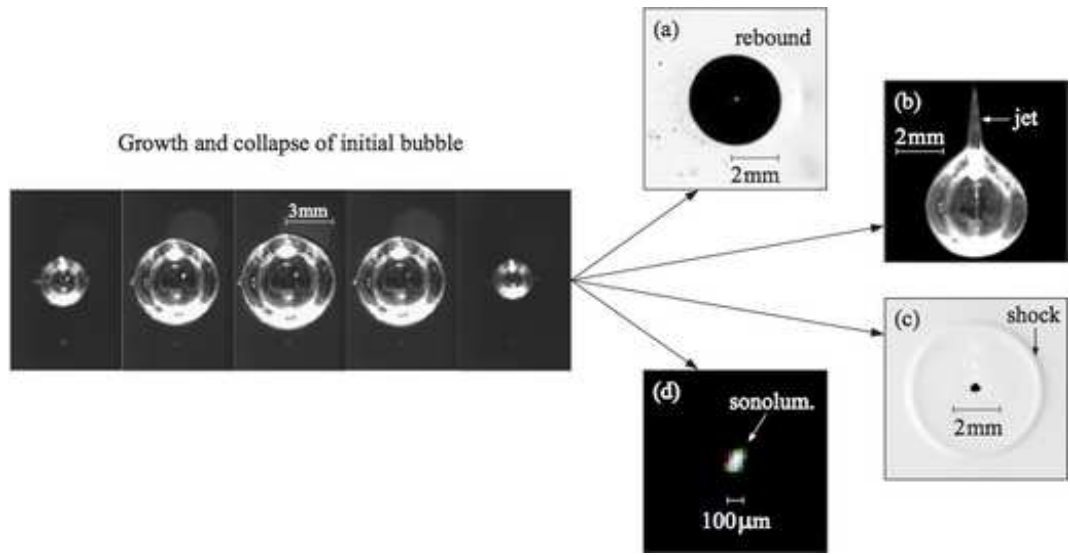}
  \caption{Illustration of the four classes of phenomena triggered by collapsing cavitation bubbles. All images were obtained using the setup presented in this paper, but they correspond to different experimental conditions. Not all phenomena are active in every collapse. For example, a spherical collapse in microgravity produces no jets.}
  \label{fig_roadmap}
\end{figure*}

All research on cavitation has departed from the ideal model of a perfectly spherical bubble collapsing within a liquid medium. The first time-solutions for the case of an empty bubble in an infinite, inviscid, incompressible liquid \citep{Stokes1847,Rayleigh1917} exhibit infinite velocities of the bubble wall at the collapse point. This branch point singularity \citep{Obreschkow2012a} highlights the insufficiency of the Stokes-Rayleigh approach, and resulted in ever more detailed considerations of the complex phenomena triggered during the bubble collapse. Those post-collapse processes can be grouped into four classes (see Fig.~\ref{fig_roadmap}):
\begin{enumerate}[(a)]
\item \textit{rebound bubbles} arising when bubbles bounce off their enclosed gas \citep{Akhatov2001};
\item \textit{micro-jets} emerging in an asymmetric collapse \citep[e.g.,][]{Blake1999,Katz1999,Wang2010,Obreschkow2006,Kobel2009};
\item \textit{shock waves} caused by the liquid compression at the collapse point \citep[e.g.,][]{Ohl1999,Schnerr2008,Obreschkow2011a};
\item \textit{thermal effects}, e.g., heating/cooling \citep{Hickling1965}, luminescence \citep[e.g.,][]{Brenner2002}, and chemical reactions \citep{Suslick1990}.
\end{enumerate}

In the wealth of established facts on those post-collapse phenomena (see review by \citealp{Lauterborn2010}), we might still be missing a unifying picture. Most prominently, a general theory for the relative importance of the different phenomena in various experimental conditions remains to be uncovered. How does the energy partition between rebound, shock, jet, and luminescence vary as a function of the liquid pressure, the liquid temperature, the gas content and the bubble sphericity? Quantitative answers to these questions promise to become an outstanding tool for optimizing virtually any application that relies on a particular feature of collapsing cavitation bubbles.

On the experimental level, three requirements must be met in the quest for a unified theory of the energy relocation of collapsing bubbles:
\begin{enumerate}
\item The setup must allow the generation of highly spherical bubbles that conserve their sphericity in the rebound, hence suppressing any jets. Jet formation can then be stimulated by adding controlled asymmetries. Only through such a precise control can different energy channels be disentangled.
\item The setup must be equipped with sensors measuring the essential phenomena: a high-speed camera to capture rebound bubbles and potential jets, a color-sensitive light sensor to measure the intensity and temperature of sonoluminescent radiation, and a time-resolved pressure sensor to capture shocks.
\item The experiment must be performed for a wide array of initial conditions to sample the parameter space covered by typical applications.
\end{enumerate}

This paper presents an experiment meeting those three criteria, and it provides online access to a selection of experimental data. Section \ref{section_experiment} describes the experimental setup, its peculiar features, and the conditions in which the experiment was performed. Section \ref{section_results} then presents results that quantify the quality of the data and illustrate the wealth of observed phenomena. For example, we present high-speed visualizations of the gravity-driven jet produced during the collapse of a spherical cavitation bubble in water subjected to normal gravity. Systematic scientific results regarding the jet formation and the energy partition between rebound and shock were presented in separate manuscripts \citep{Obreschkow2011b,Tinguely2012}. Section \ref{section_data} explains the data structure and access to the online data. Section \ref{section_conclusion} concludes the paper.

\begin{figure*}[t]
\centering
  \includegraphics[width=\textwidth]{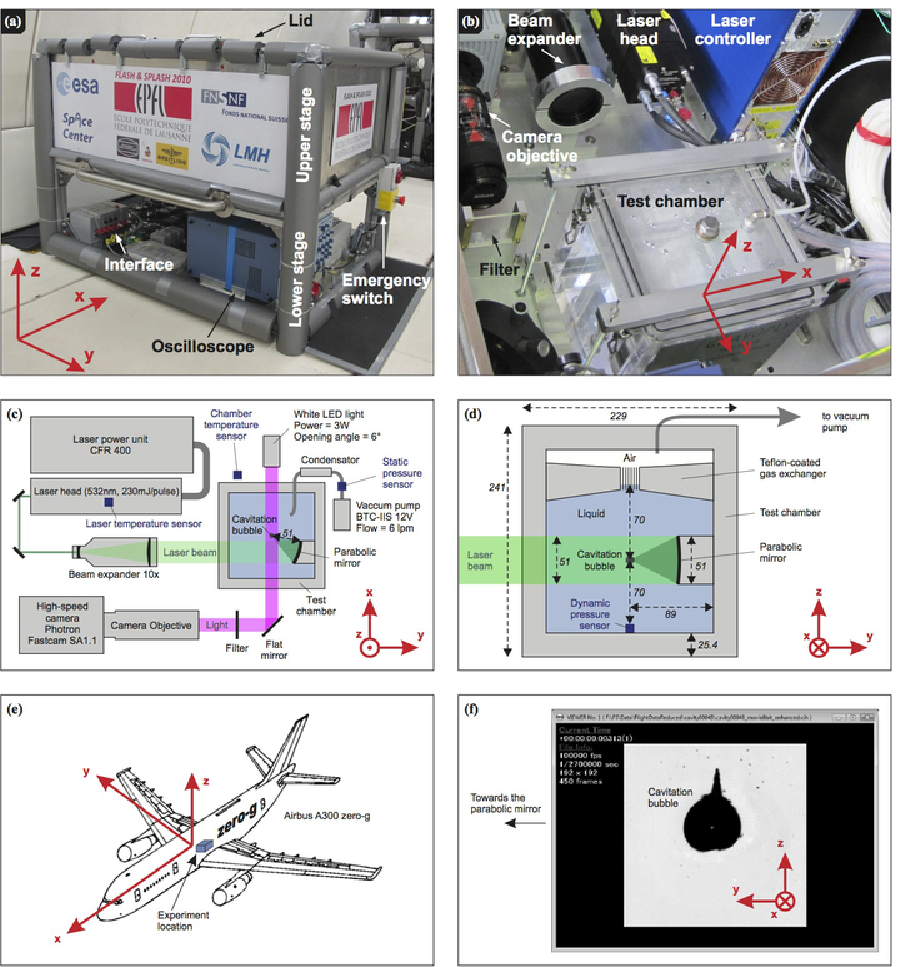}
  \caption{(Color online) Overview of the experimental setup installed inside the Airbus A300 zero-g (53rd ESA parabolic flight campaign). (a) Photo of the entire mechanical structure (``rack'') fixed inside the aircraft. The lower stage of the rack hosts electronic control/acquisition systems and power supplies. The upper stage of the rack consists of a closed box containing the test chamber, the laser-system for cavity generation, cameras and other sensors. (b) Photo of the subsystems located inside the upper stage of the rack. (c) Schematic top-view of the upper stage of the rack. (d) Side-view of the vacuum vessel, which constitutes the test chamber for the cavitation bubbles. (e) Schema of the aircraft showing the location of the experiment. (f) Image of the cavitation bubble as seen by the high-speed camera. (ALL) All six panels show the three basis vectors (x,y,z) of the Cartesian coordinate system of the aircraft. By definition, this basis is `left-handed'. Note, however, that on the camera image (e), the coordinates appear `right-handed' due to the mirror between the test chamber and the camera objective.}
  \label{fig_schema}
\end{figure*}

\section{Experimental setup}\label{section_experiment}
The experimental setup can generate a spherical vapor bubble ($\rm radius=0\!-\!7~mm$) in the middle of a cubic liquid volume ($178\times178\times150\rm~mm^3$). The growth, collapse, luminescence, and rebound of this bubble are filmed using a high-speed camera, while shock waves are recorded by a pressure sensor. To modulate and annihilate the gravity-induced pressure gradient in the liquid, the experiment is performed inside an aircraft performing parabolic flights. Section \ref{subsection_overview} overviews the mechanical structure of the setup, whose subsystems are described in Sections \ref{subsection_vessel}--\ref{subsection_sensors}. The flight manoeuvres are explained in Section \ref{subsection_flights}, and the time-sequence of a single experiment is detailed in Section \ref{subsection_trigger}. Additional background information on the experimental setup is provided by \citep{Tinguely2013}.

\subsection{Structure of the experimental setup}\label{subsection_overview}

\fig{schema} illustrates the experimental setup. The entire setup is contained within a mechanical structure called ``the rack'' (see \fig{schema}a). Its robust skeleton of strut profiles ($50\times50\rm~mm^2$ and $30\times30\rm~mm^2$) and two horizontal aluminium plates (thickness $10\rm~mm$) ensures that precision-parts, such as optical components, displace less than $\rm10~\mu m$, when passing from normal gravity to weightlessness and hypergravity, respectively. Ultimately, the whole rack withstands accelerations up to $9g$ as required for flight security. The rack is fixed inside the aircraft ``A300 zero-g'' close to the aircraft's center of gravity (\fig{schema}e), where the best level of microgravity can be achieved (see Section \ref{subsection_flights}). Globally, the rack consists of two parts: an open ``lower stage'' and a closed ``upper stage'', which can be opened via a lid. A laptop is fixed on top of this lid. A custom-designed LabView script running on this computer controls the experimental conditions, records the sensor data, and automatically triggers the bubble generation at the preselected level of gravity.

The upper stage of the rack is schematically displayed in Fig.~\ref{fig_schema}c and detailed in Figs.~\ref{fig_schema}b, d. This stage contains three main systems: a pressure-controlled test chamber filled with liquid (Section \ref{subsection_vessel}), a laser-system to generate a cavitation bubble inside the test chamber (Section \ref{subsection_bubble_generator}), and a high-speed imaging system to record the evolution of the cavitation bubble (Section \ref{subsection_camera}). Additionally, the upper stage also hosts various additional sensors, described in Section \ref{subsection_sensors}.

The lower stage of the rack hosts auxiliary electronics and power supplies. A USB-interface (PhidgetInterfaceKit 8/8/8 \#1018) converts analog sensor signals (Section \ref{subsection_sensors}) into computer-readable digital data. This interface also permits the computer to release a laser pulse for bubble generation (Section \ref{subsection_bubble_generator}), to control the pressure pump (Section \ref{subsection_vessel}), and to activate the illumination system (Section \ref{subsection_camera}). An oscilloscope records the time-variable signal of the dynamic pressure sensor (Section \ref{subsection_sensors}). All components are powered via adapters fed by a \mbox{220 V} \mbox{(50 Hz)} supply, secured with a Ground Fault Interrupter (GFI), an emergency switch, a master fast fuse (5~A), and individual fuses for each component.

\subsection{Pressure-regulated test chamber}\label{subsection_vessel}
The test chamber (see Figs.~\ref{fig_schema}b, d) is the heart of the experiment. It contains demineralized water, where cavitation bubbles are produced. This test chamber is a customized version of ``Vacuum Chamber C'' distributed by \textit{Terra Universal} and has two essential properties: (1) it is made of acrylic glass, which is optically transparent as required by the laser-based bubble generator and the high-speed camera system; (2) it can hold a vacuum, i.~e.~withstand pressures \mbox{100 kPa} \mbox{(1~bar)} below the outside-pressure, as required for a systematic study of cavitation bubbles in different pressure environments. The lid of the test chamber can be removed to access its inside and exchange the liquid.

The test chamber is filled to 80\% with liquid and to 20\% with air. This `air' also includes the vapor of the liquid at saturation pressure, as well as potential traces of laser-generated and bubble-generated gases. The air is separated from the liquid by a passive `gas exchanger' (see \fig{schema}d). In normal gravity and hyper-gravity (anti-parallel to the $z$-axis), this gas exchanger works as follows. The conical shape at the bottom of the exchanger forces gas bubbles in the liquid to migrate towards the center of the exchanger, from where they escape to the air phase across a series of vertical tubes. This permanent escape channel for gas is particularly important to constantly remove traces of laser-generated and bubble-generated gases from the liquid. Vice versa, the conical shape at the top of the gas exchanger implies that liquid trapped in the air phase flows down to the center of the exchanger, from where it leaks down to the liquid through the vertical tubes. The efficiency of this phase separation is enhanced by a hydrophobic Teflon coating, which prevents small drops and air bubbles to remain attached to the gas exchanger by surface tension. In microgravity, the small diameter-to-length ratio of the vertical tubes ($\rm diameter=4~mm$, $\rm length=40~mm$) imposes enough friction to prevent the liquid from migrating to the air phase under the effect of random gravity-fluctuations at the 0.01$g$-level. This key feature was verified experimentally by analyzing the high-speed movies and complementary webcam images (Section \ref{subsection_sensors}). No unwanted liquid/air mixing was found, except for in the few cases ($<2\%$), where the microgravity phase was perturbed by flight-turbulences causing an upwards gravitational acceleration of $a_z>0.1g$.

The chamber pressure is lowered below the ambient pressure (i.~e.~the aircraft cabin pressure of 82~kPa) using a low-power vacuum pump (Parker BTC-IIS mi\-ni\-pump). The suction of this pump is connected to the air phase of the chamber via a valve. The minimal achievable air pressure $p_{\rm air}$ in the chamber is 9~kPa. This pressure is constantly measured by a static pressure sensor (see Section \ref{subsection_sensors}), which covers the whole accessible range from 82~kPa down to 9~kPa at an RMS-precision of $\sim\!0.2\rm~kPa$ or 1\% (whichever is larger). Both the pump and the pressure sensor are connected to the computer, allowing the latter to regulate $p_{\rm air}$. Under the effect of gravity, the pressure $p_0$ inside the liquid at the position, where the cavitation bubble is generated, differs from $p_{\rm air}$ due to the weight of the water above the bubble center. Explicitly,
\begin{equation}\label{eq_pressure_correction}
	p_0 = p_{\rm air}-\rho\,a_{\rm z}\,H,
\end{equation}
where $\rho$ is the density of the liquid, $a_{\rm z}$ is the gravitational acceleration along the $z$-axis ($a_{\rm z}=-g$ in normal gravity), and $H=70\rm~mm$ is the height of the water between the bubble and the free surface of the liquid.

\subsection{Bubble generation system}\label{subsection_bubble_generator}

A single cavitation bubble is generated at the center of the test chamber by a focused laser pulse. This pulse of 8~ns duration, 532~nm wavelength, and selectable energies up to 230~mJ emerges from a frequency-doubled Q-switched Nd:YAG-laser (Quantel CFR 400). We first describe the physics behind laser-based bubble generators and subsequently elaborate on the particular advances incorporated in the present experiment.

Laser-based techniques for generating bubbles have been widely explored \citep{Lauterborn1972,Lauterborn1975,Tomita1990,Vogel1996,Philipp1998,Vogel1999,Ohl1999,Wolfrum2003,Brujan2002,Byun2004,Ohl2006,
Lim2010}. They generally work as follows. A parallel pulsed laser beam, typically of a few nano-seconds duration and visible color, is focussed inside an optically transparent liquid medium -- say water. Despite the transparency of the water, enough energy is absorbed in the focal point that a small liquid volume ($\rm radius\lesssim0.1~mm$) heats up to ionizing temperatures (i.~e.~several 1000~K). This liquid volume hence transforms into a plasma \citep{Vogel1996,Vogel1999,Byun2004} yielding an enormous pressure of roughly 0.5~GPa per 1000~K assuming an ideal gas law. The hot plasma expands explosively, first by compressing the liquid in its immediate environment, thus generating a spherical shock wave, then by radially accelerating the liquid. The newly formed bubble typically expands to a radius $R_0$ 100-times larger than that of the original plasma. This enormous expansion (volume factor of $10^6$), as well as the efficient heat transfer through radiation and conduction, quickly cool the expanding plasma, which hence deionizes and then recondensates. Apart from non-condensible gases generated by the laser pulse \citep[e.g., H$_2$ and O$_2$, see][]{Sato2013}, the bubble has `forgotten' its thermal history by the time it reaches its maximal radius. Therefore, a vapor bubble generated by a laser pulse is for many practical purposes identical to a cavitation bubble, i.~e.~a bubble that grows due to low liquid pressure rather than high gas pressure.

Since the plasma pressure is \textit{per se} isotropic, laser-generated bubbles expand similarly in all directions, resulting in spherical bubbles. However, deviations from a sphere, can originate from the non-spherical shape of the microscopic plasma and from inhomogeneities in the liquid pressure. Both sources of asymmetry have been minimized in our experiment, as described hereafter.

Firstly, in order to obtain a highly symmetric laser-generated plasma, the parallel laser beam must be focussed into a single point at a large angle of convergence ($>30^\circ$), since such an angle reduces the heating of water inside the laser beam in the vicinity of the focal point. Lenses are not very suitable for such a focus \citep[Chapter 4 in][]{Tinguely2013}, since they suffer from monochromatic aberration at large angles of convergence and since the quality of the lens-focus depends on the wavelength of the laser and on the refractive index of the liquid, thus requiring a customized lens for each laser-liquid pair. Those issues can be bypassed by using a concave parabolic mirror to focus the parallel laser beam. Therefore, the present experiment uses for the first time a parabolic mirror (Edmund Optics, $2"\times2"$ 30 deg off-axis parabolic gold mirror, \#NT47-088), such as shown in Fig.~\ref{fig_schema}c, d. This mirror exhibits an angle of convergence of about $53^\circ$. A gold-coated mirror surface was selected to suppress a degradation by corrosion. The gold surface absorbs about 30\% of the laser energy at 532~nm -- an acceptable artefact given that the surface density of the absorbed energy remains a factor 100 below the damage threshold. To avoid that a part of the laser beam is reflected back into the laser source, which might damage the latter, we chose an off-axis mirror as show in Fig.~\ref{fig_schema}c. The parabolic mirror has a declared surface accuracy of $\lambda$/4 (RMS) and roughness of less than 17.5~nm (RMS). These high-quality specifications allow the laser to be focused in a volume smaller than the volume of the generated plasma, hence avoiding initial asymmetries of the plasma.

Secondly, pressure irregularities in the liquid must be minimized in the quest for a spherical bubble collapse. Any pressure gradient $\del p$ in the vicinity of the collapsing bubble can cause the formation of a jet directed against $\del p$ \citep{Obreschkow2011b}. If the liquid is at rest before the bubble is generated, then $\del p$ can have two origins: a static uniform pressure gradient $\del p=\rho\a$ is due to a (gravitational) acceleration $\a$, and a dynamic non-uniform $\del p$ is due to the interaction between the moving bubble surface and nearby boundaries. In fact, \cite{Blake1988} showed that bubbles collapsing near a flat rigid boundary form a jet directed \textit{towards} that boundary, whereas bubbles collapsing near a flat free surface form a jet directed \textit{away} from that surface. This boundary-induced jet-formation dominates over gravity-induced jet-formation, if (square of Eq.~8.8 in Ref.~\citealp{Blake1988})
\begin{equation}\label{eq_blake}
	\lambda\equiv \frac{h^2 \rho a}{\rmax\deltap}<0.2,
\end{equation}
where $h$ is the distance between the bubble center and the flat boundary (rigid or free surface), $a$ is the norm of the gravitational acceleration, $\rmax$ is the maximal bubble radius, and $\deltap$ is the pressure difference between $p_0$ and the pressure inside the bubble (approximately the vapour pressure of the liquid). In most past experiments $\lambda$ is much smaller than 0.2, either intentionally or due to the difficulty of generating a bubble far away from components necessary for the bubble generation (e.~g.~optical lenses). Hence the effects of the gravity-induced pressure gradients normally remain obscured by boundary effects. To decrease these effects, such that $\lambda>0.2$, the value of $h$ in Eq.~(\ref{eq_blake}) must be maximized. To do so, we chose a test chamber that is large compared to the cavitation bubbles (dimensions in Fig.~\ref{fig_schema}d) and a parabolic mirror with a large focal distance of 51~mm. Given the off-axis geometry of this mirror, the average distance between the focal point and the mirror is $h=\rm55~mm$. Given water ($\rho=10^3\rm~kg~m^{-3}$) in normal gravity ($a=g=9.81\rm~m~s^{-2}$), Eq.~(\ref{eq_blake}) then implies that gravity-driven jet-formation occurs if $\rmax\deltap\lesssim150\rm~kg~s^{-2}$. In standard units this condition for `gravity-domination' can be rewritten as
\begin{equation}\label{eq_gravity_jet}
	\left(\frac{\rmax}{\rm mm}\right)\left(\frac{\deltap}{\rm bar}\right)<1.5.
\end{equation}
For example, at a pressure $\deltap=0.2\rm~bar$ all bubbles with maximal radii $\rmax$ below 7.5~mm will be dominantly deformed by gravity, rather than by boundary effects.

In summary, the generation of a spherical bubble requires a parabolic mirror with a large angle of convergence to maximize the symmetry of the initial plasma and large focal length to minimize the effects of nearby boundaries. Together these two requirements imply a large diameter of the parabolic mirror (here 51~mm). A Galilean beam expander (Newport High-Energy Laser Beam Expander HB-10X) is used to expand the parallel laser beam to this large diameter (see Fig.~\ref{fig_schema}b, c).

The energy $E_{\rm p}$ of the laser pulse can be adjusted between $\sim0\rm~mJ$ and $\sim200\rm~mJ$ by varying the time-delay between laser-pumping and Q-switching from $500\rm~\mu s$ down to $170\rm~\mu s$. Only a minor fraction of this energy is transformed into the cavitation bubble, while other parts are converted into the shock-wave at the plasma formation, converted into heat of the liquid, absorbed by the chamber wall and gold mirror, or simply transmitted across the fluid. The maximal bubble energies $E_{\rm b}$ that can be obtained with this setup are about $E_{\rm b}\approx12\rm~mJ$. These energies are calculated from the observed maximal bubble radius via
\begin{equation}\label{eq_bubble_energy}
	E_{\rm b} = \frac{4\pi}{3}\rmax^3\deltap,
\end{equation}
where
\begin{equation}\label{eq_deltap}
	\deltap = p_0-p_{\rm v}.
\end{equation}
$p_0$ is the water pressure, calculated via Eq.~\ref{eq_pressure_correction} from the measured pressure $p_{\rm air}$, and $p_{\rm v}$ is the vapor pressure calculated from the measured water temperature via Antoine's equation \citep{Antoine1888}.

\subsection{High-speed visualization system}\label{subsection_camera}

\begin{figure}[b]
\centering
  \includegraphics[width=\columnwidth]{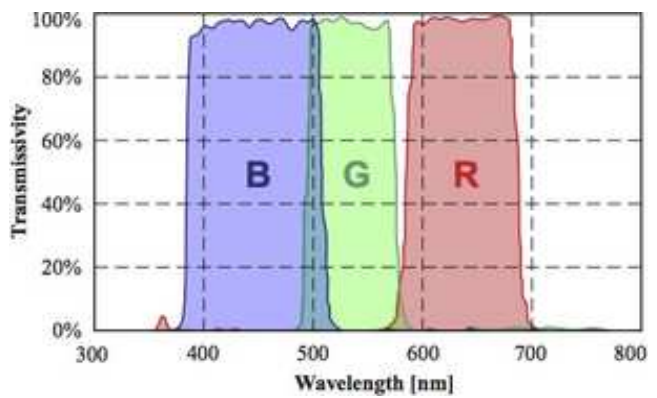}
  \caption{(Color online) Transmissivity spectra of the optional color filters placed in front of the high-speed camera. The spectral properties and planarity of these filters satisfy the high standards of astronomical photography (Ref.: Baader RGBC-CCD $65\times65\times3$ mm Filters).}
  \label{fig_filters}
\end{figure}

The visualization system contained within the upper stage of the rack is schematically represented in Fig.~\ref{fig_schema}c. It consists of a high-speed camera (Photron Fastcam SA1.1) fitted with a 135~mm objective (Nikon Zoom-Micro 70--180mm f/4.5--5.6D) and exchangeable astrophysical filters to analyze the sonoluminescent flash. We use a set of three broad-band filters (Baader Anti-reflective RGB CCD-fil\-ters, $65\times65$~mm) to analyze the sonoluminescent flash in three distinct wavebands (R, G, B), according to the transmission spectra shown in Fig.~\ref{fig_filters}. An additional clear filter is used in order to maintain the same focus when no color filter is applied. This RGB-filter system proofed to be useful for an approximate characterization of the luminescence temperature.

The camera operates at inter-frame times down to $2\rm~\mu s$ and exposure times of 370~ns. The spatial resolution of the data released with this paper is $69.0\rm~\mu m/pixel$ horizontally ($y$-axis) and $68.7\rm~\mu m/pixel$ vertically ($z$-axis). The typical field-of-view is $17.7\times17.6\rm~mm^2$ ($256\times256\rm~pixels$), although it can be expanded to $70.7\times70.3\rm~mm^2$ ($1024\times1024\rm~pixels$).

A parallel LED background illumination (StageLine 3W LED-36Spot) provides a white background, against which cavitation bubbles and jets appear as black absorption features, such as illustrated in Figs.~\ref{fig_roadmap}a and \ref{fig_schema}f. This background-illumination exhibits the advantage that the bubble surface is defined very clearly. Hence, the dimensions of the bubbles and jets can be measured reliably at a sub-pixel resolutions of $7\rm~\mu m$. Using the minimal exposure times of 370~ns, this illumination also allows the visualization of shock waves in shadowgraphy as shown in Fig.~\ref{fig_roadmap}c. An alternative front illumination (not yet available aboard the parabolic flights), also allows us to visualize bubbles as bright features in front of a dark background. This illumination is more appropriate to study the 3-dimensional morphology and the inner structure of the bubbles, such as the thin vertical jet visible inside the rebound bubble in Fig.~\ref{fig_roadmap}b.

No illumination is used when studying the sonoluminescent flash from the collapsing bubble itself. The CMOS-sensor of the high-speed camera is sufficiently sensitive to detect this faint flash on most single bubble collapses, even when filters are used. Section \ref{subsection_phaseIII} explains how the signal-to-noise ratio of this data can be increased by exploiting the cross-correlation between the sonoluminescent pulse and the collapse shock recor\-ded by the dynamic pressure sensor (Section \ref{subsection_sensors}).

The high-speed camera generates a significant power of heat, worth about 100~W, which must be continually removed from the closed upper stage of the rack through an active air-cooling system. 

\subsection{Sensors}\label{subsection_sensors}

In addition to the high-speed camera an array of sensors is used for experimental and security reasons. These sensors are described one-by-one in the following.

\subsubsection{Sensors for science and control}\label{subsubsection_science_sensors}

\begin{figure}[t]
\centering
  \includegraphics[width=\columnwidth]{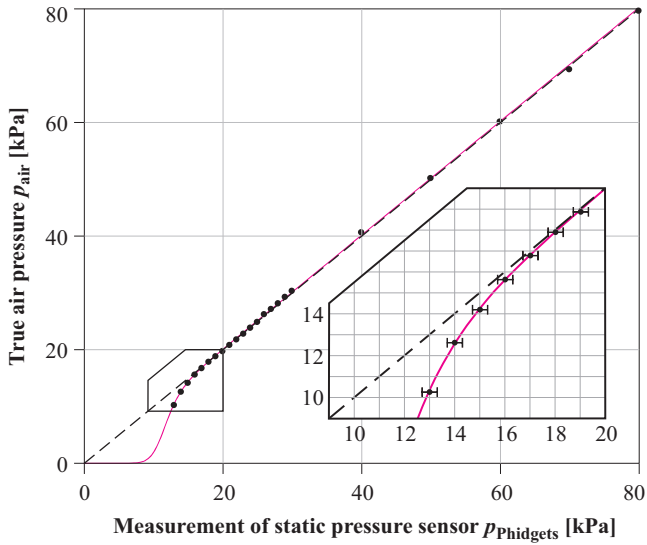}
  \caption{(Color online) Non-linear calibration used to extend the nominal pressure range of the static pressure sensor (20--250\rm~kPa) down to 9~kPa. The data points show the true air pressure $p_{\rm air}$ applied to the static pressure sensor against the measurements $p_{\rm Phidgets}$ of the latter. The non-linear function (solid line) of Eq.~(\ref{eq_pressure_calibration}) provides an excellent fit to these data.}
  \label{fig_pressure_calibration}
\end{figure}

\begin{figure}[t]
\centering
  \includegraphics[width=\columnwidth]{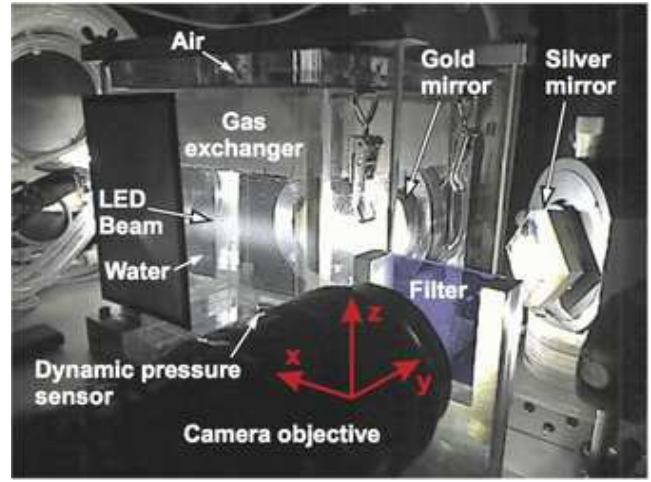}
  \caption{(Color online) Webcam image allowing a safe monitoring of the experiment when the box is closed during laser operation. The components described in Fig.~\ref{fig_schema} that are visible in the Webcam image are labelled.}
  \label{fig_webcam}
\end{figure}

\begin{figure*}[t]
\centering
  \includegraphics[width=0.9\textwidth]{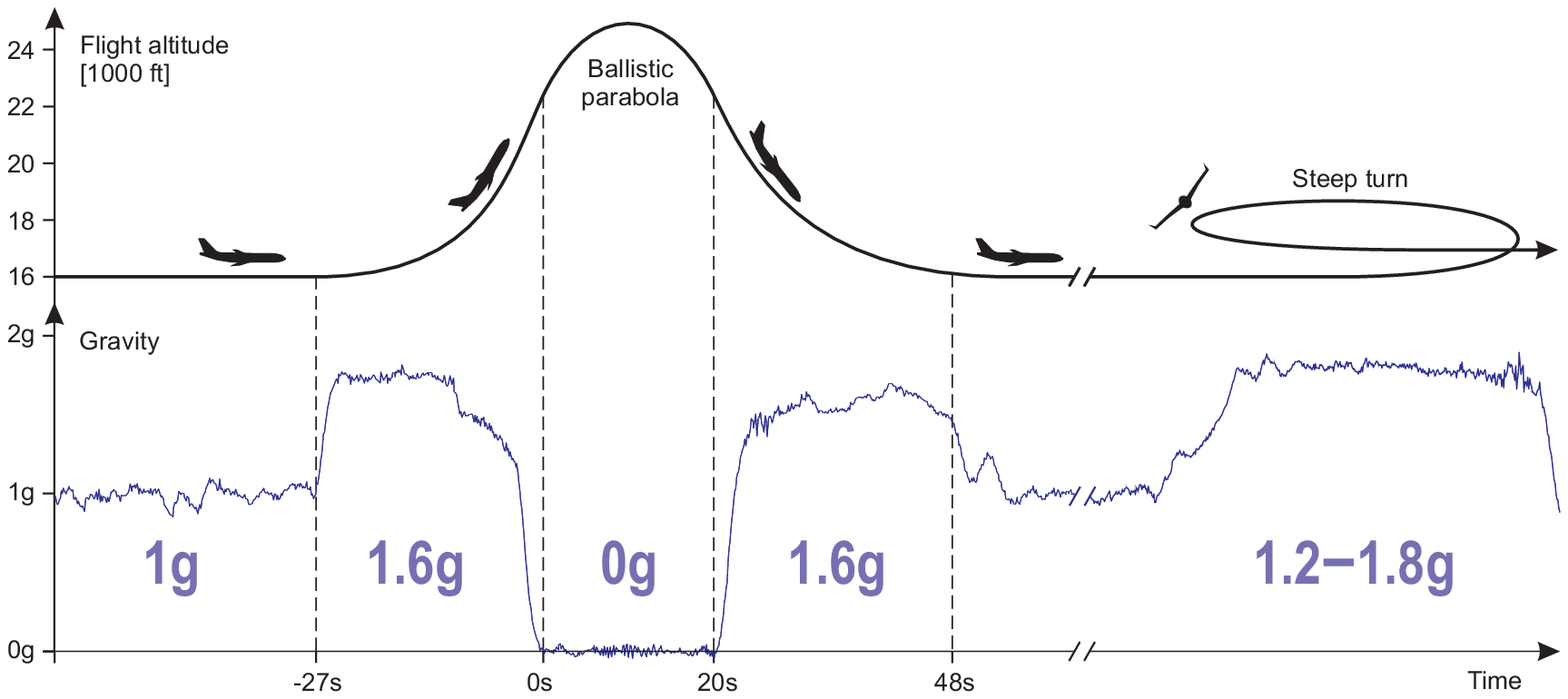}
  \caption{(Color online) Illustration of a parabola and a steep turn flown by the aircraft A300 zero-g. A representative measurement of the vertical acceleration $-a_{\rm z}$ as a function of time is plotted underneath the flight trajectory.}
  \label{fig_flight}
\end{figure*}

\begin{itemize}
\item[\sym] \textit{Dynamic pressure sensor:} A high-frequency pressure sensor is used to record the transit of the shock waves produced at the initial plasma generation and at the subsequent collapses of the principal cavitation bubble and the cascade of rebound bubbles. Our first experiments with already published results \citep{Obreschkow2011b,Obreschkow2012a,Tinguely2012} use a custom-designed piezo-resistive pressure sensor described in Section 4.2.1 of \citep{Hasmatuchi2011}. This sensor has a baseband bandwidth (constant gain to 25~kHz, resonant at 100~kHz) and a pressure range (linear in \mbox{0--500~kPa}), which suffice to accurately measure the timing of the shock passage, but the shock front cannot be sampled in detail. Measurements of the shock energy therefore require extrapolation techniques, which proofed to be useful but rather uncertain \citep{Tinguely2012}. To improve these measurements the pressure sensor was replaced by a modern hydrophone (1.0~mm needle with 28~$\mu$m sensor, manufactured by \textit{Precision Acoustics PAL}). The bandwidth of this hydrophone extends to 12~MHz, hence allowing a detailed sampling of the shock front, although the actual peak-pressure, typically on the order of hundreds of MPa, remains inaccessible. The pressure signal, represented by an electrical voltage is pre-amplified and recorded by an oscilloscope (Le\-Croy WaveRunner 6050A, 500 MHz bandwidth).
\item[\sym] \textit{Static pressure sensor:} A gas pressure sensor (Phidgets Gas Pressure Sensor, \#1115), connected to the chamber-side of the vacuum pump (Fig.~\ref{fig_schema}c), monitors the pressure of the air inside the test chamber at an RMS-precision of 0.2~kPa or 1\% of the measured pressure (whichever is larger). Using the non-linear calibration displayed in Fig.~\ref{fig_pressure_calibration}, the nominal linear pressure range of (20--250)~kPa, could be extended down to 9~kPa without loss of precision. For a measured static pressure $p_{\rm Phidgets}$ the true air pressure is calculated by
\begin{equation}\label{eq_pressure_calibration}
	p_{\rm air} = \frac{p_{\rm Phidgets}}{1+\left(\frac{p_{\rm Phidgets}}{11.5\rm~kPa}\right)^{-11}}.
\end{equation}
\item[\sym] \textit{Chamber temperature sensor:} A thermometer is used to measure the water temperature, from which the vapor pressure is calculated. Our first experiments \citep{Obreschkow2011b,Tinguely2012} used a sensor (Phidgets Precision Temperature Sensor, \#1124) attached to the outside wall of the test chamber (Fig.~\ref{fig_schema}c). This sensor has a statistical precision of 0.5~K, but since it is not in direct contact with the water, the measurements are subject to systematic uncertainties and time-delays. To improve the measurements the sensor was subsequently replaced by a water-resistant thermistor (USB Thermistor DTU6024C-004-S), inserted into the water near a corner of the test chamber. This sensor reliably measures the water temperature at a statistical accuracy of 0.1~K.
\item[\sym] \textit{Accelerometer:} A 3-axis accelerometer (Phidgets Accelerometer 3-Axis, \#1059), fixed on the rack, re\-cords the inertial acceleration $\a$ (including gravity) in the range $[-3g,+3g]=[-29.4\rm~m\,s^{-2},+29.4~m\,s^{-2}]$ in each dimension, at an RMS-precision of 0.0019$g$ ($a_{\rm x}$ and $a_{\rm z}$) and 0.0029$g$ ($a_{\rm y}$). The bandwith of 30~Hz lies safely above the fastest g-jitter of the aircraft around 10~Hz. This accelerometer is used both to trigger the generation of a cavitation bubble at the desired level of gravity and for the post-processing, if we require a gravity-measurement ra\-ther than a nominal value associated with the particular flight maneuver. An example of a measurement of $a_{\rm z}$ as a function of time is shown in Section \ref{subsection_flights} (Fig.~\ref{fig_flight}).
\item[\sym] \textit{Webcam:} A USB-webcam (Logitech Quickcam Pro, $960\times720\rm~pixels$, 30 frames per second) is installed inside the upper stage of the experimental rack in order to safely monitor the setup, while the experiment is running. All major components of the upper stage are visible on the webcam image, such as shown in Fig.~\ref{fig_webcam}.
\item[\sym] \textit{Manual trigger:} A hand-hold button, stowed in the lower stage of the rack, can be used to release an experimental cycle (bubble generation and high-speed imaging). This button serves as a backup in case of an in-flight failure of the automatic trigger.
\end{itemize}

\subsubsection{Sensors for experimental security}\label{subsubsection_security_sensors}

\begin{itemize}
\item[\sym] \textit{Mechanical lid sensors:} Two electro-mechanical contact sensors, directly connected to the ``interlock'' connector of the laser power unit, ensure that no laser pulse can ever be fired while the lid of the upper rack stage is open.
\item[\sym] \textit{Magnetic lid sensor:} As an additional security measure, a magnetic contact sensor (Phidgets Miniature Magnetic Contact Switch BR-2023), connected directly to the computer, prevents the latter from commanding the laser, if the lid is open.
\item[\sym] \textit{Light sensors:} Two light sensors (Phidgets Precision Light Sensor \#1127) monitor the illuminance inside the upper rack stage at the 1~lux-level. If traces of light are still measured, while the magnetic lid sensor is closed and the LED illumination system is turned off, the laser is stopped automatically. This mechanism provides an additional level of security, excluding the possibility that the laser is fired if the upper rack stage is damaged in such away that light can escape even though the lid is closed.
\item[\sym] \textit{Laser temperature sensors:} A temperature sensor (Phidgets Precision Temperature Sensor, \#1124) attached to the laser head (Fig.~\ref{fig_schema}c) and four thermometers built into the laser-box by the manufacturer control the laser temperature at all times. If any of these sensors indicates a dangerous temperature, the laser is stopped immediately.
\item[\sym] \textit{Amperemeter:} An inductive amperemeter (Phidgets 30 Amp Current Sensor AC/DC, \#1122) monitors the current drawn by the experiment at 1.5\% accuracy. Any current irregularities below the 5~A threshold of the master fuse, are displayed as warnings on the computer.
\end{itemize}

\subsection{Parabolic flights}\label{subsection_flights}

The experiment is performed on parabolic flights flown by the aircraft ``A300 zero-g'' -- so far during the 53rd (year 2010) and 56th (year 2012) parabolic flight campaign of the European Space Agency (ESA). The typical layout of a flight campaign is as follows. Over three consecutive flight days, the aircraft performs $3\times31=93$ parabolas, offering about 20~s of microgravity each. Depending on demand, the steep turns are added to generate stable levels of hyper-gravity at $1.2g$, $1.4g$, $1.6g$, and $1.8g$. A single parabola and a steep turn at $1.8g$ are illustrated in Fig.~\ref{fig_flight} together with a representative measurement of the vertical acceleration, recorded by the accelerometer.

The parabolas are flown in a ballistic manner, i.~e.~at a constant horizontal speed, thus imitating the cen\-ter-of-gravity of motion of a free-falling body in the absence of air drag. By doing so, the aircraft experiences weightlessness in the same way as a spacecraft in orbit. There are, however, small gra\-vity-perturbations due to tiny deviations in the aircraft's trajectory. This residual g-jitter exhibits an RMS-amplitude of $0.01g$ at frequencies of 1--10~Hz. Furthermore, the slow pitching of the aircraft causes a small centrifugal acceleration that is strongest at the nose and at the tail of the aircraft, where it reaches about $0.01g$. Our experiment has an allocated space near the aircraft's center of gravity to avoid this source of systematic perturbation.

\subsection{Trigger sequence}\label{subsection_trigger}

Each experimental cycle with a single cavitation bubble follows the same sequence. A software trigger activates the laser unit as soon as the user-defined level of gravity is reached and stabilized for 1~s. This unit initiates the pumping of the laser and uses an internal precision clock to actuate the Q-switch at the preset time-delay. When the Q-switch is activated, thus releasing the laser pulse, a signal simultaneously triggers the high-speed camera and the oscilloscope recording the dynamic pressure. We chose to synchronize the camera and oscilloscope with the Q-switch to ensure that they are always synchronized with the cavitation bubble, independently of the Q-switch delay.

\section{Results}\label{section_results}

This section describes the data gathered specifically during the 53rd parabolic flight campaign conducted in October 2010. We first overview the sampled parameter space (Section \ref{subsection_parameter_space}) and illustrate the life-cycle of a single cavitation bubble (Section \ref{subsection_typical_bubble}). Thereafter, we successively address technical details associated with different stages in this life-cycle (Sections \ref{subsection_phaseI}--\ref{subsection_phaseIII}). Note that systematic scientific analyses are published separately \citep[][and forthcoming]{Obreschkow2011b,Obreschkow2012a,Tinguely2012}.

\subsection{Parameter space overview}\label{subsection_parameter_space}

\begin{figure}[t]
\centering
  \includegraphics[width=\columnwidth]{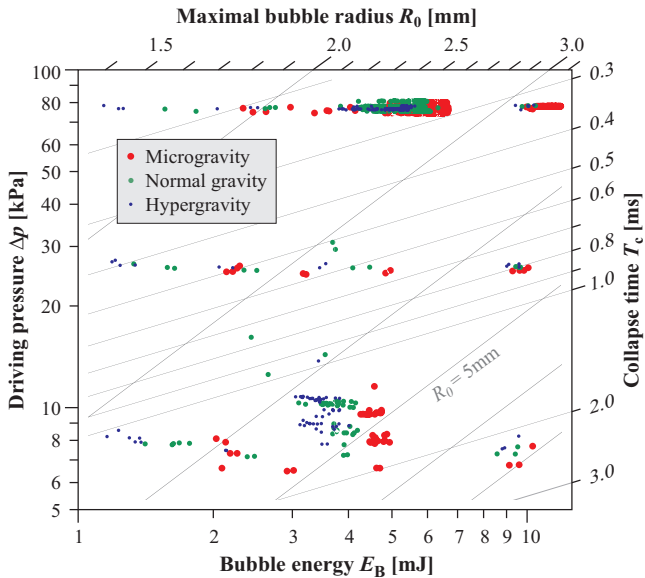}
  \caption{(Color online) Overview of the bubble parameters probed on parabolic flights. The data is binned in three gravity bins (colors) according to the nominal gravity level of the flight manoeuvre. The four parameters $E_{\rm b}$, $\deltap$, $\rmax$, and $\tc$ are related via Eqs.~(\ref{eq_bubble_energy}) and (\ref{eq_rayleigh_time}) as demonstrated in Section \ref{subsection_phaseII}. Therefore those four parameters can be reduced to only two dimensions, as done in this plot.}
  \label{fig_parameter_space}
\end{figure}

A total of 4887 single cavitation bubbles were generated during the flight maneuvers, using demineralized water at room temperature. Of those bubbles, 247 were observed with background illumination, while the remaining 4640 bubbles were recorded without illumination to detect the sonoluminescent flash (details in Section \ref{subsection_phaseIII}). Three parameters could be varied independently:
\begin{itemize}
\item[\sym] The energy of the laser was varied to modulate the bubble energy $E_{\rm b}$ between 1.1~mJ and 11.9~mJ.
\item[\sym] The air pressure $p_{\rm air}$ of the chamber was adjusted to vary $\deltap$ between 6.5~kPa and 81.2~kPa.
\item[\sym] The gravitational acceleration $a_{\rm z}$ slowly alters between $0g$ and $1.8g$ during the flights, causing the pressure gradient $|\del p|$ to vary between $0$ and $1.8\rho g$.
\end{itemize}
As shown in Section \ref{subsection_phaseII}, the evolution of our bubbles is closely described by Rayleigh's theory \citep{Rayleigh1917}. Therefore, given $E_{\rm b}$ and $\deltap$, the maximal bubble radius $\rmax$ ($1.5-7.0\rm~mm$) and its collapse time $\tc$ ($157-2439\rm~\mu s$) are uniquely determined via Eq.~(\ref{eq_bubble_energy}) and
\begin{equation}\label{eq_rayleigh_time}
	\tc = f\rmax\sqrt{\frac{\rho}{\deltap}},
\end{equation}
where the Rayleigh-factor is $f\approx0.91468$. Using Eqs.~(\ref{eq_bubble_energy}) and (\ref{eq_rayleigh_time}), any two of the variables $\{E_{\rm b}, \deltap, \rmax, \tc\}$ fully determine the remaining two. In logarithmic space these relations are linear, i.e.,
\begin{eqnarray}
	\log E_{\rm b} & = & k_1 + 3\log\rmax + \log\deltap \\
	\log \tc & = & k_2 + \log\rmax - 0.5\log\deltap
\end{eqnarray}
with constants $k_1\equiv\log(4\pi/3)$ and $k_2\equiv\log f+0.5\log\rho$. Using these relations we can represent all four parameters $\{E_{\rm b}, \deltap, \rmax, \tc\}$ in a two-dimensional logarithmic plot as shown in Fig.~\ref{fig_parameter_space}. The situations probed by our experiment are represented as points, whose color indicates the level of gravity.

\subsection{Overview of a bubble's life}\label{subsection_typical_bubble}

\begin{figure*}[t]
\centering
  \includegraphics[width=\textwidth]{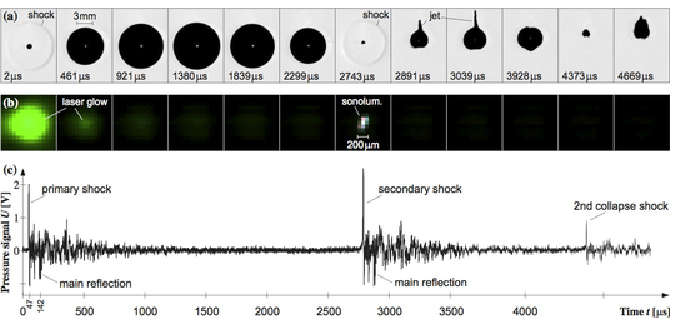}
  \caption{(Color online) Illustration of the time-resolved measurements characterizing the evolution of single cavitation bubbles. Fig.~\ref{fig_single_bubble}a shows selected 370~ns-exposures of a high-speed movie of a bubble evolving in a static pressure $\Delta p=8.98\pm0.03\rm~kPa$ and a gravitational acceleration $|\a|=(1.847\pm0.005)g$. This bubble attains a maximal radius $\rmax=4.549\pm0.007\rm~mm$ and exhibits a collapse time of $\tc=1410.83^{+0.19}_{-0.02}\rm~\mu s$. The background-illumination used in this visualization turns the bubble and shocks into dark absorption features. Fig.~\ref{fig_single_bubble}b shows a color-movie obtained without illumination, in order to visualize the light emitted by the bubble itself (luminescence). Each color channel (R, G, B) of this movie is the average of about 800 single movies obtained by using the color filters (Fig.~\ref{fig_filters}) on 2400 single cavitation bubbles. Fig.~\ref{fig_single_bubble}c displays the voltage recorded by the dynamic piezo-pressure sensor (Section \ref{subsection_sensors}) during the evolution of the bubble shown in Fig.~\ref{fig_single_bubble}a. The physical phenomena revealed in these figures are explained in Section \ref{subsection_typical_bubble}.}
  \label{fig_single_bubble}
\end{figure*}

Fig.~\ref{fig_single_bubble} illustrates the time-resolved measurements characterizing the evolution of a single bubble. 

Fig.~\ref{fig_single_bubble}a shows the high-speed movie of a bubble specified in the caption of Fig.~\ref{fig_single_bubble}. This bubble is generated at the time $t=\rm0~\mu s$. At $t=\rm2~\mu s$ the ``primary'' shock, driven by the expanding plasma, can be seen about 3~mm from the bubble center. This shock is visible due to the density-dependence of the water's refractive index \citep{Cho2001}. The apparent thickness of the shock of $0.5\pm0.1\rm~mm$ can be explained by the smearing of a sound wave ($c\approx1480\rm~m\,s^{-1}$) traveling for the exposure time of 370~ns. Hence, the intrinsic shock thickness must be smaller than $0.1\rm~mm$. On the interval $0\rm~\mu s-2741\rm~\mu s$ the bubble grows and collapses. At the collapse point, a ``secondary'' shock wave is emitted. The collapse is followed by a rebound, during which a gravity-driven jet emerges from the bubble.

Fig.~\ref{fig_single_bubble}b shows a synthetic 3-color movie based on 2400 individual movies of bubbles similar to that shown in Fig.~\ref{fig_single_bubble}a, but filmed without background-illumination. Those movies were taken while placing the three color filters (R, G, B) successively in front of the camera, such that each color band yields about 800 movies. Those movies, all taken at inter-frame spacings and exposure times of $66.7\rm~\mu s$ (15~kHz), were then rescaled in time to match the bubble collapse time of Fig.~\ref{fig_single_bubble}a and averaged to obtain a noise-reduced signal. The resulting color movie clearly shows the white sonoluminescent radiation at the collapse point, which indicates a temperature on the order of $10^4\rm~K$ in the black body approximation. The beginning of the movie shows a bright green spot, which is an inevitable after-glow of the intense green (532~nm) laser pulse.

Fig.~\ref{fig_single_bubble}c shows the dynamic pressure signal. At $t\approx\rm47~\mu s$, the plasma-shock appears on the dynamic pressure signal, the delay being due to the distance of 70~mm between the shock center and the sensor (Fig.~\ref{fig_schema}d). At $t\approx\rm142~\mu s$, the pressure exhibits a negative peak, associated with the stratification wave generated when the shock is reflected at the free water surface. The pressure signal also indicates significant shocks at the first and second bubble collapse. In addition to these clear features in the pressure signal, the signal also shows various spurious oscillations, partially associated with higher-order reflections, partially caused by the sensor's eigenmodes around and beyond 100~kHz.

The different stages of the bubble evolution will now be discussed in detail over the Sections \ref{subsection_phaseI} to \ref{subsection_phaseIII}.

\subsection{Phase I: Bubble generation}\label{subsection_phaseI}

\begin{figure}[t]
\centering
  \includegraphics[width=\columnwidth]{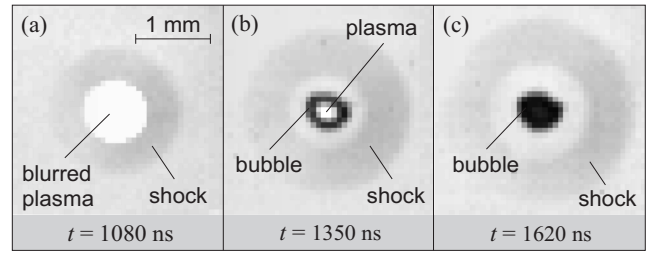}
  \caption{Interleaved high-speed sequence of the bubble generation from a hot, laser-generated plasma. The laser pulse at $t=0$ lasts for 8~ns. (Bubble energy $E_{\rm b}\approx3\rm~mJ$, driving pressure $\deltap\approx20\rm~kPa$, exposure times are 370~ns)}
  \label{fig_interleaving}
\end{figure}

The bubbles are generated by a 8~ns laser pulse focussed inside the liquid (Section~\ref{subsection_bubble_generator}). Fig.~\ref{fig_interleaving} shows three interleaved photo\-graphs of the initial bubble growth at an inter-frame spacing of only 270~ns. These photographs were extracted from three different movies of three similar bubbles (bubble energy $E_{\rm b}\approx3\rm~mJ$, driving pressure $\deltap\approx20\rm~kPa$). The exact timing of the images was determined from the timing of the primary shock on the dynamic pressure sensor. In this way it is possible to mimic inter-frame spacings much shorter than the minimal inter-frame spacing of $1.5\rm~\mu s$.

Fig.~\ref{fig_interleaving}a unveils that a spherical shock wave is generated while the laser-heated region shines at visible wavelengths. The brightness of this light-emitting region causes the camera sensor to saturate and blur. The white disk at the center of Fig.~\ref{fig_interleaving}a is therefore much larger than the actual light-emitting region. The existence of radiation at visible wavelengths witnesses temperatures of several 1000~K, hence the light-emitting medium must be ionized. On Fig.~\ref{fig_interleaving}b this plasma has cooled down to temperatures that still emit visible light, without, however, saturating the camera sensor. At this stage, a dark proto-bubble centered about the plasma becomes visible. Stated differently, Fig.~\ref{fig_interleaving}b demonstrates that there is a moment in the bubble-generation process, where the bubble has already formed and the shock has already detached from the bubble, although the bubble gas is still hot enough to be partially ionized. However, only 270~ns later, i.~e.~at about 0.1\% of the Rayleigh time (Eq.~\ref{eq_rayleigh_time}), the plasma has fully recombined (Fig.~\ref{fig_interleaving}c), leaving a bubble filled with water vapor and minor amounts of other gases \citep{Sato2013}.

\subsection{Phase II: Bubble growth and collapse}\label{subsection_phaseII}

\begin{figure}[t]
\centering
  \includegraphics[width=\columnwidth]{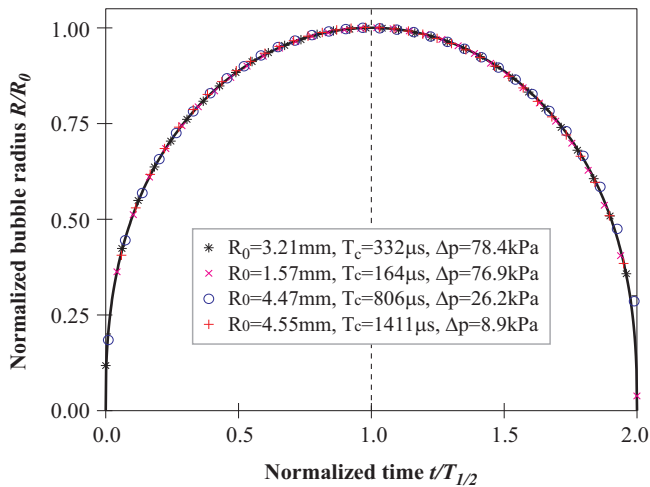}
  \caption{(Color online) Growth and collapse motion of four distinct cavitation bubbles, represented in normalized coordinates. The solid line is the prediction of the Rayleigh model.}
  \label{fig_rayleigh_fit}
\end{figure}

\begin{figure}[t]
\centering
  \includegraphics[width=\columnwidth]{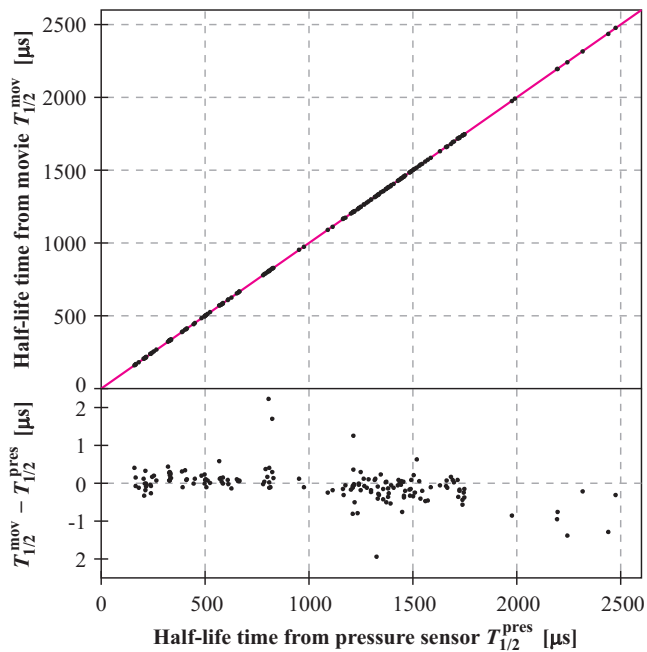}
  \caption{(Color online) Comparison of the half-life times $\tl$ of 163 cavitation bubbles measured using the dynamic pressure sensor ($\tl^{\rm pres}$) and the high-speed movie ($\tl^{\rm mov}$), respectively. The close agreement between these independent measurements demonstrates the small measurement uncertainties of both types of measurements.}
  \label{fig_collapse_times}
\end{figure}

Once generated, the bubbles grow and collapse. Fig.~\ref{fig_rayleigh_fit} shows this evolution in terms of the bubble radius $R(t)$, normalized to the maximal radius $\rmax$ and the total life-time $T_0$. Although these data correspond to four different experimental conditions, the normalized evolutions are congruent and closely approximated by the Rayleigh model (solid line in Fig.~\ref{fig_rayleigh_fit}). In fact, we have recently shown that the collapse motion of some of our bubbles fits the Rayleigh model within less than 0.1\% in terms of radius \citep{Obreschkow2012a}. This close agreement with the Rayleigh model suggests that the effects of surface tension and viscosity are negligible -- a feature attributed to the large ($R_0>1\rm~mm$) size of the bubbles considered in this work.

Fig.~\ref{fig_rayleigh_fit} also reveals that $R(t)$ is nearly symmetric in time. Hence the collapse time $\tc$, i.e, the duration from $R=R_0$ to $R=0$, is well approximated by the half-life time $\tl\equiv T_0/2$. Slight asymmetries in the time-evolution are physically interesting, but lie beyond the technical description envisaged in this paper. Instead, we shall now show that the experiment can measure $\tl$ with a relative uncertainty below $10^{-3}$. In general, $\tl$ can be measured independently from the pressure signals and high-speed movies. The pressure-based approach measures the time-difference between the primary shock, emitted at the bubble generation, and the secondary shock, emitted at the final stage of the bubble collapse. Since those two shocks imprint a similar response on the sensor (see Fig.~\ref{fig_single_bubble}c), their separation in time can be very accurately determined using an auto-correlation applied to the spline-interpolated voltage-signal. Alternatively, $\tl$ can also be measured directly from the high-speed movies. These movies use different inter-frame spacings from $1.5\rm~\mu s$ to $20\rm~\mu s$. By measuring the bubble radii on each frame at sub-pixel resolution and applying a Rayleigh-interpolation to estimate the continuous time-evolution of the radius, $\tl$ can be measured with an uncertainty much smaller than the inter-frame times.

We can then compare the half-life times $\tl^{\rm pres}$ and $\tl^{\rm mov}$ extracted from the pressure signals and movies, respectively. To this end we use all 163 cavitation bubbles that have high-speed movies and full pressure data. These bubbles cover half-life times of $\tl=157-2439\rm~\mu s$, energies of $E_{\rm b}=1-10\rm~mJ$, and pressures of $\deltap=9-80\rm~kPa$. The comparison (Fig.~\ref{fig_collapse_times}) reveals that $\tl^{\rm pres}$ and $\tl^{\rm mov}$ systematically agree within a standard deviation of
\begin{equation}\label{eq_rms_tl_diff}
	\sqrt{\left\langle\left(\tl^{\rm mov}-\tl^{\rm pres}\right)^2\right\rangle} \approx 380\rm~ns.
\end{equation}
Assuming that the measurement uncertainties of $\tl^{\rm pres}$ and $\tl^{\rm mov}$ are uncorrelated, this result implies that $\tl^{\rm pres}$ and $\tl^{\rm mov}$ both have a statistical measurement uncertainty below $380\rm~ns$. In other words, both the pressure sensor and the high-speed camera measure $\tl$ at a relative precision between 0.01\% and 0.1\%, depending on the actual value of $\tl$. This is an extreme accuracy compared to the nominal time-resolution of the pressure sensor ($\sim10~\mu s$ at resonance) and the inter-frame spacing ($1.5-20\rm~\mu s$) of the movies, thus demonstrating the power of using smart time-interpolations. Since those interpolations are simpler in the case of $\tl^{\rm pres}$ than $\tl^{\rm mov}$, we hereafter set $\tl=\tl^{\rm pres}$, whenever a reliable pressure signal is available, and $\tl=\tl^{\rm mov}$ otherwise.

\begin{figure}[t]
\centering
  \includegraphics[width=\columnwidth]{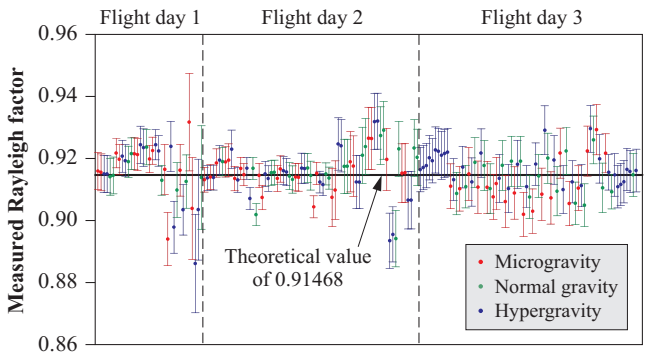}
  \caption{(Color online) Measured Rayleigh factors $f$, defined in Eq.~\ref{eq_rayleigh_time}, compared against the theoretical value.}
  \label{fig_rayleigh_factor}
\end{figure}

Upon assuming $\tc=\tl$ and provided parallel measurements of $\rmax$ from the high-speed movies, we can then check if the pairs $\{\tc,R_0\}$ are systematically consistent with the Rayleigh model. To do so, we solve Eq.~(\ref{eq_rayleigh_time}) for the Rayleigh factor $f$ while propagating the measurement uncertainties of $\tc$, $R_0$, and $\deltap$. The values of $f$ hence obtained are consistent with the theoretical value (Fig.~\ref{fig_rayleigh_factor}). Gravity has no obvious effect on the life-time of the cavitation bubbles besides the indirect effect of changing $\deltap$ via Eq.~(\ref{eq_pressure_correction}).

\subsection{Phase III: Inflection point and rebound}\label{subsection_phaseIII}

The collapse of a cavitation bubble progresses until a violent increase in the gas pressure abruptly decelerates the collapse motion \citep{Akhatov2001}. The compression of the non-condensable bubble gas \citep{Tinguely2012} increases the internal energy of this gas to ionizing temperatures, thus causing a sonoluminescent flash. Simultaneously, the immense gas pressure significantly compresses the surrounding liquid, thus driving the secondary shock away from the bubble. Finally, the gas pressure stops the collapse motion and inverts it, thus causing a rebound bubble. As outlined in the introduction, the key objective of our setup is to analyze the energy partition between spherical rebound, micro-jet, shock wave, and thermal effects such as luminescence (see Fig.~\ref{fig_roadmap}). To achieve this goal, we record the evolving shape of the rebound, the pressure of the shock wave, and the sonoluminescent flash.

\begin{figure}[t]
\centering
  \includegraphics[width=\columnwidth]{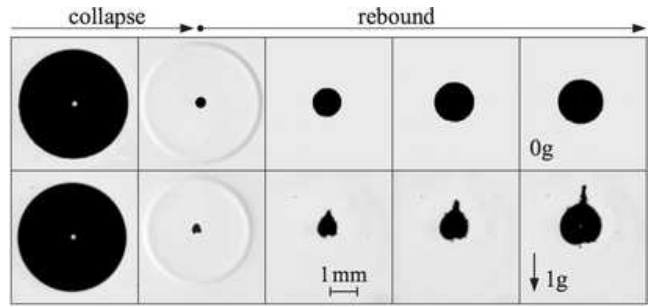}
  \caption{Collapse and rebound of two cavitation bubbles imaged against a bright background. The top row shows a bubble in microgravity, where the rebound remains spherical, while the bottom row shows a bubble in normal gravity, where the rebound is deformed by a jet propagating against the gravity vector. [adopted from \cite{Obreschkow2011b}]}
  \label{fig_jet}
\end{figure}

The evolving rebound bubble is imaged by the grey\-scale high-speed camera used with background illumination. Those visualizations reveal a gravity-driven jet (Fig.~\ref{fig_jet}), which disappears in microgravity conditions and gets enhanced in hyper-gravity. This jet is in actual fact a vapor envelop of a thin liquid micro-jet, visible in Fig.~\ref{fig_roadmap}b. These jets are caused by the gravity-induced pressure gradient$\del p=\rho\mathbf{a}$. However, since the jets cannot `know' that the pressure gradient is due to gravity, we can consider gravity as a generic tool to study the effects of any uniform pressure gradient $\del p$. Furthermore, $\del p$ is the first factor in the spatial Taylor expansion of any inhomogeneous pressure field. Hence, this study also applies, to first order, to any inhomogeneous pressure field. We recently presented detailed investigations of the jet as a function of $\del p$ and other experimental parameters \citep{Obreschkow2011b}. This study found that the volume of the vapor jet, normalized to the volume of the spherical component of the rebound, depends linearly on the non-dimensional parameter
\be\label{eq_def_zeta}
	\zeta \equiv |\del p|\rmax/\Delta p.
\ee
A theoretical model confirmed this scaling law. Ground-based follow-up experiments with viscous mixtures of water and glycerine further showed that the jet-size is independent of the viscosity of the liquid. The jet's independence of viscosity, as well as its independence of surface tension can be seen as a consequence of the global conservation law of momentum (Kelvin-impulse) during the inflection of the collapse motion \citep{Obreschkow2011b}. A different explanation for the insignificance of viscosity and surface tension is that their effects are masked by inertial forces, which increase more rapidly as $R(t)\rightarrow0$, according to the Rayleigh-Plesset equation \citep{Plesset1971}.

\begin{figure}[b]
\centering
  \includegraphics[width=\columnwidth]{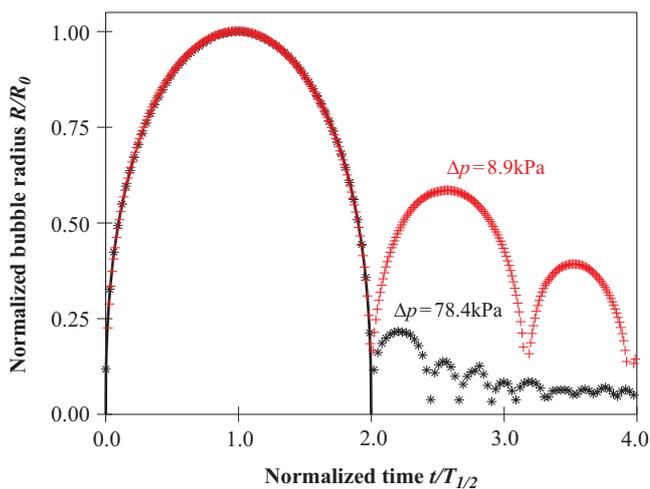}
  \caption{(Color online) Two bubbles with different normalized rebound radii. The fraction of the initial bubble energy recovered in the rebound bubble is related to the non-dimensional parameter $\xi$, defined in Eq.~(\ref{eq_def_xi}). These two bubbles have values of $\xi\approx600$ (black $\ast$) and $\xi\approx70$ (red +), the difference originating from different values of $\deltap$. The other bubble parameters are given in the caption of Fig.~\ref{fig_rayleigh_fit}.}
  \label{fig_rebound}
\end{figure}

The shock wave is registered on the dynamic pressure sensor with a time-delay of $47~\mu\rm s$. Although the piezo-resistive sensor used on the 53rd parabolic flight campaign is unable to sample the wavefront, the integrated response of this sensor, $\int{\rm d}t~s^2(t)$, can be used to estimate the energy of the shock. In recent work \citep{Tinguely2012}, we calibrated this method and analyzed systematic variations of the energy ratio between the shock wave and the rebound bubble. This empirical study, backed-up with a first-order model for liquid shocks, suggests that the energy partition between shock and rebound is fully determined by the single non-dimensional parameter
\be\label{eq_def_xi}
	\xi \equiv \frac{\gamma^6\deltap}{p_{\rm g}^{1/\gamma}(\rho c^2)^{1-1/\gamma}},
\ee
where $p_{\rm g}$ is the pressure of the non-condensable gas at the maximal bubble radius, $\gamma\approx1.3$ is the adiabatic index of the non-condensable gas, $\rho$ is the liquid density, and $c$ is the liquid's speed of sound. An example of two bubbles with different values of $\xi$ is shown in Fig.~\ref{fig_rebound}.

Finally, the greyscale high-speed camera can capture the sonoluminescent flash instead of imaging the rebound bubble, if the illumination is turned off. The luminescence only ever appears on a single frame of the movie. Hence this luminescence lasts much shorter than the minimal inter-frame spacing of about $1.5\rm~\mu s$, in agreement with detailed theoretical and experimental studies suggesting typical durations of less than $10\rm~ns$ \citep{Lauterborn2010}. Our setup can image the luminescence either directly, through any of the three color filters (RGB, see Fig.~\ref{fig_filters}), or through a new spectral analyzer not yet available on the 53rd parabolic flight campaign. Only one filter can be used at a time; hence three-color studies must rely on several bubbles produced under identical conditions and observed successively in the R, G, and B band. The problem with this method is the limited reproducibility of the luminescence. In fact, we found that bubbles of identical energy and driving pressure $\deltap$, thus identical radius and life-time, vary in their luminescence brightness about 10-times more than expected from pure shot noise considerations. These brightness fluctuations are probably related to the microscopic size of the sonoluminescent plasma, which makes it highly sensitive to minor perturbations and easily obscured by nuclei and impurities in the water. Any robust result must therefore rely on statistical averages of many single bubbles. For this reason, 20-times more experiments were conducted to measure luminescence than to image the rebound bubble on the 53rd parabolic flight campaign. Corresponding scientific results will be presented in forthcoming work.

\section{Online data}\label{section_data}

Most data presented in this paper can be accessed online at \textsl{bubbles.epfl.ch/data}. So far, this online data contains 247 cavitation bubbles imaged against a white background during the 53rd parabolic flight campaign. These bubbles have been numbered from `cavity00001' to `cavity00247' in chronological order. Each bubble has a single zip-file, containing up to eight data files: a log-file with header data, a file with the dynamic pressure data, and six files for the high-speed movies.  Some bubbles are missing either the pressure file of the movie files due to experimental difficulties in-flight.

\subsection{Log-files (cavity00000\_log.txt)}

The log-files are ascii-files describing the experimental conditions of the bubble in all available detail. An example of a log-file is given in Tab.~\ref{tab_log_file}. In particular, the log-files contain all available measurements of the static sensors. Furthermore, they contain comments on problems with certain sensors, as well as potential comments made by the experimenter. At the end of each log-file, we list basic post-processing results, including the maximal bubble radius $\rmax$, the collapse time $\tc$, as well as the driving pressure $\deltap$. Given those values and their measurement uncertainties, we also provide the most likely values of $\rmax$, $\tc$, and $\deltap$ assuming that they satisfy the Rayleigh model in Eq.~(\ref{eq_rayleigh_time}). Those values, forced to satisfy Eq.~(\ref{eq_rayleigh_time}), are called reference values. Explicitly, they have been obtained by maximizing the total probability function $\rho(\rmax)\rho(\tc=f\rmax\sqrt{\rho/\deltap})\rho(\deltap)$, where $\rho(\rmax)$, $\rho(\tc)$, and $\rho(\deltap)$ denote the individual probability distributions associated with the individual measurements of $\rmax$, $\tc$, and $\deltap$, respectively. Most experimental values are provided with uncertainty intervals. They represent standard deviations where the noise is considered Gaussian, and smallest intervals containing 67\% of the events, where the noise is non-Gaussian.

\subsection{Pressure files (cavity00000\_pressure.txt)}

The pressure files are ascii-files containing the pre-amp\-lified voltages of the piezo-resistive dynamic pressor sensor, sampled at 2.5~MHz using an oscilloscope (Le\-Croy WaveRunner 6050A, 500 MHz bandwidth). Each file contains 20,000 data points, hence covering a time-interval of 8~ms, sufficient to capture the primary and secondary shock of all bubbles. The data is organized in two columns. The first column gives the time in s relative to the instant of the laser pulse that generates the bubble. The shock impacts the pressure sensor with a time-delay of about $47~\rm\mu s$, due to the 70~mm distance between the bubble center and the pressure sensor. The second column gives the electrical voltages recorded by the oscilloscope. These voltages have been calibrated to $13.55\rm~\mu V~Pa^{-1}$. However, the sensor's response is too slow and too limited in dynamic range to measure the actual shock pressure. The pressure signal can nonetheless be used to measure the life-time of the bubble and to estimate the shock energy as outlined in Section \ref{section_results}.

Some bubbles exhibit dynamic pressure data without clear shocks and/or clear low-frequency oscillations on the order of 1~kHz. Those data are affected by spurious air bubbles sitting on top of the pressure sensitive piezo-element. Typically, the data deteriorated in this way yield values below 5 in the `Dynamic pressure signal-to-noise' entry of the log-files.

\subsection{Movie files (cavity00000\_movieX)}

There are up to six movie files per cavity specified by the placeholder `X'. They are three actual video-files and three corresponding header-files, readable by the software \textit{Photron Fastcam Viewer (PFV)}, provided by the manufacturer of the high-speed camera. The three files are the original 12-bit greyscale video (cavity00000\_movie12bit.mraw), readable by PFV, as well as two standard 8-bit greyscale movies in avi-format, readable by most video software. The first avi-file (cavity00000\_movie8bit.avi) is a simple resampling of the original 12-bit scale. For most applications it is virtually identical to the original 12-bit movie, but generally more easy to process. The second avi-file (cavity00000\_movie8bit\_enhanced.avi) is graphically en\-han\-ced movie with a flattened background brightness and reduced noise achieved via dark frame subtraction.

\section{Summary and future prospects}\label{section_conclusion}

In this work, we have presented an experimental setup able to produce highly spherical cavitation bubbles in conditions so rigorously isotropic that the pressure gradient of normal gravity is the most important source of asymmetry. In fact, the experiment can clearly visualize a gravity-driven jet, propagating against the vector of gravitational acceleration. In the past, such observations of gravity-driven jets could only be obtained in the case of large bubbles ($\rmax\!>\!1\rm\,cm$) collapsing in hyper-gravity \citep{Benjamin1966}, hence requiring less spherical initial conditions. Given that normal gravity is the most important anisotropy of our bubbles, we then conducted the whole experiment in microgravity aboard parabolic flights. This setting thus allowed us to generate bubbles more spherical than theoretically possible in any ground-based experiment.

As argued in the introduction, this experiment offers an ideal platform to uncover a general model for the energy partition between rebound, jet, shock, and thermal effects during the collapse of a single cavitation bubble. Such a model would make a crucial tool for reducing cavitation damage and optimizing beneficial applications of cavitation. While this paper is focussed mostly on technical aspects, we have also illustrated two important scientific results: The strength of the micro-jet is proportional to the non-dimensional parameter $\zeta$ of Eq.~(\ref{eq_def_zeta}) \citep{Obreschkow2011b}, and the energy partition between the rebound and the shock depends monotonically on the non-dimensional parameter $\xi$ of Eq.~(\ref{eq_def_xi}) \citep{Tinguely2012}.

In making some data obtained during the 53rd para\-bolic flight campaign (October 2010) available online, we open a possibility for other research groups to work with exclusive microgravity data. Moreover, next generation data obtained on the 56th parabolic flight campaign (May 2012) and forthcoming ones will allow a more detailed view of the luminescence. Ultimately, this research is aimed at unifying the quickly diversifying field of cavitation research in providing a global model for the energy distribution between all cavitation related effects.

\small{

{\renewcommand{\arraystretch}{1.5}
\onecolumn
\begin{longtable}{p{4.5cm}p{8.4cm}p{3.23cm}}\label{tab_log_file}

\textbf{Name} & \textbf{Additional explanation} & \textbf{Value} \\ [1ex] \hline \\ [-2.5ex]
\endfirsthead

\textbf{Name} & \textbf{Additional explanation} & \textbf{Value} \\ [1ex] \hline \\ [-2.5ex]
\endhead

Unique cavity ID & Unique 5-digit identifier of the bubble; the bubbles presented here range from 00001 to 04887. & 00147 \\ 
Version of data reduction & Version of the code used to reduce the movie- and pressure-files and to produce the log-files. & 2.0 \\
Date of cavity generation & -- & 21-Oct-2010 \\
Time of cavity generation [UTC+2h] & -- & 09:59:13.995 \\
Flight maneuver & `dayX' shows the flight day $\rm X=1,2,3$; `paraYY' means that the bubble was produced in modified gravity during the parabola $\rm YY=00,..,30$ or in the normal gravity succeeding that parabola; `turnZZ' means that the bubble was produced in the hyper-gravity of the steep turn $\rm ZZ=01,..,08$. & day3turn04 \\
Flight maneuver sub-index & Bubbles belonging to the same `Flight maneuver' are distinguished with an increasing sub-index, ranging from 1 to 200. & 3 \\
Problems found in data inspection & -- & none \\
High-speed movie available & Is `false' if the movie is not available due to technical problems in-flight. & true \\
Control sensors available & Is `false' if data from the control sensors (e.~g.~acceleration, static pressure, temperatures) is not available due to technical problems in-flight. & true \\
Dynamic pressure available & Is `false' if data from the dynamic pressure sensor is not available due to technical problems in-flight. & true \\
Dynamic pressure signal-to-noise & Ratio between the peak voltage and the RMS-noise of the dynamic pressure sensor. Values below 10 indicate that the pressure signal is deteriorated by gas on the piezo-sensor. & 18.2 \\
Pulse rate & `single' means that two successive bubbles are spaced by several seconds, such that any interference between those bubbles can be ignored; `7Hz' means that bubbles were generated at a 7~Hz rate, hence gases produced by a bubble may not be completely removed until the next bubble is formed. & single \\
Pulse index & Equals `1' if the `Pulse rate' is `single'; and counts from `1' upwards if the `Pulse rate' is `7Hz'. & 1 \\
Dynamic pressure mode & If `DC', the pressure signal is recorded without filter, if `AC', the pressure signal has been processed by a high-pass filter. Note that the offset tension is negative in `DC' and zero in `AC'. \\
Q-switch delay [us] & Delay between laser pumping and Q-switching. Larger values indicate less laser energy. Range: 170 (corresponding to $\sim\rm200~mJ$ pulses) to 500 (vanishing pulse energy). & 275 \\
Nominal vessel pressure [kPa] & Target value of the pressure $p_{\rm air}$, which the computer tries to achieve; only use this value, if no value is available for `Measured vessel pressure'. & 10.3 \\
Measured vessel pressure [kPa] & Pressure $p_{\rm air}$ measured by the static pressure sensor. & 10.32$\pm$0.15 \\
Derived pressure at cavity level [kPa] & Pressure $p_0$ at the bubble center before bubble generation, obtained via Eq.~(\ref{eq_pressure_correction}). & 11.58$\pm$0.15 \\
Vapor pressure of the water [kPa] & Vapor pressure $p_{\rm v}$ derived from a measurement of the water temperature at the beginning of each flight day. & 2.46 \\
Measured ambient pressure [kPa] & Pressure in the aircraft cabin. & 83.00$\pm$0.5 \\
Nominal gravity $\mid$g$\mid$ [9.8m/s\textasciicircum 2] & Norm of the gravitational acceleration that the aircraft tried to achieve during the respective flight maneuver. Only use this values if no `Measured gravity' is available. & 1.8 \\
Measured gravity $\mid$g$\mid$ [9.8m/s\textasciicircum 2] & Measured norm $a=|\a|$ of the gravitational acceleration. & 1.847$\pm$0.005 \\
Measured gravity g\_x [9.8m/s\textasciicircum 2] & Measured gravitational acceleration $a_{\rm x}$. & -0.194$\pm$0.005 \\
Measured gravity g\_y [9.8m/s\textasciicircum 2] & Measured gravitational acceleration $a_{\rm y}$. & -0.017$\pm$0.005 \\
Measured gravity g\_z [9.8m/s\textasciicircum 2] & Measured gravitational acceleration $a_{\rm z}$. & -1.837$\pm$0.005 \\
Movie size & Size in pixels of the frames of the high-speed movie. & 256x256 \\
Number of frames & Number of frames in the high-speed movie. & 600 \\
Frame rate [Hz] & Frame rate of the high-speed movie. & 67500 \\
Shutter speed [s] & Exposure time of each frame. & 1/2700000 \\
Microns per pixel (horizontal) & Horizontal size of each pixel in the high-speed movie. & 69.9 \\
Microns per pixel (vertical) & Vertical size of each pixel in the high-speed movie. & 69.6 \\
LED backlight & Is `on', if the back-illumination is used (e.~g.~Fig.~\ref{fig_single_bubble}a); is `off' of no illumination is used (e.~g.~Fig.~\ref{fig_single_bubble}b). & on \\
Camera diaphragm & F-number of the camera objective, that is the ratio between the aperture diameter and the focal length. & 11 \\
Camera filter & Type of filter placed in front of the camera as shown in Fig.~\ref{fig_schema}c; `clear', if a clear filter is used that corresponds to no filter, but conserves the focus; `red', `green', or `blue' if one of the three color filters described in Fig.~\ref{fig_filters} is used. & clear \\
Frame of luminescence & If `LED backlight' is `off', this is the frame number in the high-speed movie, on which the sonoluminescent light is either clearly detected or expected from the collapse shock measured by the dynamic pressure sensor. & N/A \\
Sonoluminescent signal [summed pixel value] & Number of counts of the sonoluminescent signal. WARNING: It is necessary to check in the high-speed movies whether individual pixels are saturated (i.~e.~at value 4192 or 1.0 depending on the movie-type). & N/A \\
Measured cavity radius [mm] & Maximal radius $\rmax$ of the cavitation bubble as measured from the high-speed movie. & 4.548$\pm$0.007 \\
Measured half-life time [us] & Best measurement of the half-life time $\tl$ (half the time from the bubble generation to the first collapse). This time is measured either from the dynamic pressure sensor or the high-speed movie, if no dynamic pressure is available. & 1370.40$\pm$0.10 \\
Measured collapse time [us] & Estimated collapse time $\tc$ from the maximal bubble radius to the collapse point. This time is obtained from $\tl$ via $\tc/{\rm s}=1.107(\tl/{\rm s})^{1.011}$, an equation calibrated on a subset of 20 cavitation bubbles imaged with increase spatial and temporal resolution. & 1410.91$\pm$0.10 \\
Source of measured collapse time & (see above) & Dynamic pressure sensor \\
Measured delta pressure [kPa] & Pressure $\Delta p$ calculated via Eq.~(\ref{eq_deltap}). & 9.12$\pm$0.15 \\
Reference cavity radius [mm] & Most likely value of the maximal bubble radius $\rmax$, given the measurements of $\rmax$, $\tc$, and $\Delta p$, and imposing Eq.~(\ref{eq_rayleigh_time}). & 4.551+0.007-0.007\\
Reference collapse time [us] & Most likely value of the collapse time $\tc$, given the measurements of $\rmax$, $\tc$, and $\Delta p$, and imposing Eq.~(\ref{eq_rayleigh_time}). & 1410.81+0.20-0.00 \\
Reference delta pressure [kPa] & Most likely value of the driving pressure $\Delta p $, given the measurements of $\rmax$, $\tc$, and $\Delta p$, and imposing Eq.~(\ref{eq_rayleigh_time}). & 8.70+0.03-0.03 \\
Reference energy of cavitation bubble [mJ] & Value $E_{\rm b}$ calculated via Eq.~(\ref{eq_bubble_energy}), given the reference values of $\rmax$ and $\Delta p$. & 3.44+0.03-0.03 \\
\hline
\caption{Explanation of the log-files; example of \textsl{cavity00147\_log.txt}.}
\label{tab_log_file}
\end{longtable}
\twocolumn}
}

\bibliographystyle{spbasic}      

\end{document}